\documentclass[preprint,showpacs,prc,aps,showkeys]{revtex4-1}
\usepackage{graphicx}% Include figure files
\usepackage{adjustbox}%adjust table size
\usepackage{dcolumn}% Align table columns on decimal point
\usepackage{bm}% bold math
\usepackage{hyperref}% add hypertext capabilities
\usepackage{amsmath}
\usepackage{subfig}
\usepackage{tabularx}
\begin{document}

\preprint{PLB}

\title{Systematics of dynamic moment of inertia in super-deformed bands in Mass $\sim$150 region}

\author{S. Roy}
\affiliation{Department of Nuclear and Atomic Physics, 
Tata Institute of Fundamental Research, Mumbai- 400 005, India}

\begin{abstract}
An empirical semi-classical model have been proposed to investigate the nature of  
dynamic moment-of-inertia , of the super-deformed (SD) bands in 
nuclei of mass 150 region. The model incorporates an additional frequency 
dependent distortion, to the dynamic moment-of-inertia term akin to a vibrational 
component to explain the extreme spin structure of these bands. Using this model 
two separate components to the dynamic moment of inertia, $\Im^{(2)}$ have been  
identified for the SD band structure for the mass 150 region. Three distinct nature 
of the moment-of-inertia, also have been identified using the two parameter model.  
\begin{description}

\item[PACS numbers] 
\date{\today}

\end{description}
\end{abstract}

\maketitle
\date{\today}

\section{Introduction}

The super-deformation in atomic nuclei has been an exhaustively studied intriguing 
high spin phenomenon. Till date, more than three hundred super-deformed (SD) 
bands have been reported across the mass regions~\cite{singh2002}. The 
super-deformation in a nucleus have been characterized by an extended ellipsoidal 
shape, that may be imagined as a rugby ball. For the ideal SD case the 
ellipsoidal long axis is twice the short axis in length, and corresponds to 
a quadrupole deformation, of $\beta \sim 0.6$~\cite{nolan1988}. Such an 
extended shape is stable at high spin and rotational frequency due to the 
interplay of collective and single particle degrees of freedom. The primary 
factors for the stability of shape is due to the shell corrections, which 
produces a local minima in the potential-energy-surface, and the decrease of 
the Coulomb force due to larger than average separation of protons. 
In addition, at this high spin range (I $\geq 30 \hbar$), presence of highly 
mixed $j$-states presents an interesting domain for theoretical studies. 
There has been many exhaustive theoretical studies, using mean field models;
Woods-Saxon~\cite{dudek1985}, anharmonic oscillator potential
~\cite{ragnarsson1980, bengtsson1981}, and Skyrme-Hartree-Fock~\cite{satula1996}.
Quite a few semi-classical macroscopic models are has also been successful in presenting 
a good description of the SD bands~\cite{swaitecki1987,harris1967,wood1974,wu1987,wu1992}.

In the super-deformation regime, the nucleus exhibits nearly a perfect 
quadrupole behaviour, with a very small regular separation in the rotational 
level energies. This feature may manifests as identical SD bands in 
neighbouring nuclei~\cite{cyrus1995} due to similar moment-of-inertia. 
Certain features of the moment-of-inertia has also been studied by 
the semi-classical model, where a stiff core part was identified at high spin 
(Eq.~\ref{eq1})~\cite{swaitecki1987}.

\begin{equation}
\Im=\frac{3}{4}\Im_{rigid}+\frac{1}{4}\Im_{irrotational}
\label{eq1}
\end{equation}

\noindent where, $\Im_{rigid}$ and $\Im_{irrotational}$, are the respective rigid-core
and irrotational contribution to the total moment-of-inertia, $\Im$~\cite{swaitecki1987}.

Characteristics of a rotational band can be investigated, from the energy 
of the levels (E$_L$) and the angular momentum (I). The moment-of-inertia ({$\Im$}),
kinematic-moment-of-inertia ($\Im^{(1)}$) and dynamic-moment-of-inertia ($\Im^{(2)}$),
here after also referred to as D$^{MOI}$, are the three quantities that defines the 
evolution of the band with the spin. 
These are given by,
\begin{eqnarray}
\Im^{(1)}= I\left(\frac{dE}{dI}\right)^{-1} \nonumber \\ 
\Im^{(2)}=\left(\frac{d^2E}{dI^2}\right)^{-1}
\label{eq2}
\end{eqnarray}

For an ideal rotor, $\Im \equiv \Im^{(1)} \equiv \Im^{(2)}$. However, in a realistic 
scenario, the excited levels in a nucleus have also been determined by pairing 
effect between the valance nucleons and the centrifugal stretching due to 
rotation~\cite{satula1996}. Both the effects results in a D$^{MOI}$ that is dependent on
the rotational frequency.

For a deformed nucleus in which the spheroidal collective field is cranked
about a fixed axis, a general expression for moment-of-inertia was given by,
A. Bohr, B. R. Mottelson, S. A. Moszkowski and D. R Inglis
~\cite{bohr1955,moszkowski1956,Inglis1956},

\begin{eqnarray}
\Im_{Inglis}= \frac{\hbar}{\omega_2\omega_3}\left[\frac{(\omega_2-\omega_3)^2}{\omega_2+\omega_3}
\sum(m+n+l) + \frac{(\omega_2+\omega_3)^2}{\omega_2-\omega_3}\sum(n-m)\right] 
\label{eq3}
\end{eqnarray}

\noindent where, $l$, $m$, and $n$ are the oscillator quantum number in the three 
dimensions defining the single nucleon states, $\omega_2$ is the oscillator 
frequency in $x$ and $y$ direction and $\omega_3$ is the oscillator frequency 
in the z direction. For $\omega_2\neq\omega_3$ $\sum(n-m)$ is zero. 
The first term is associated with the irrotational-flow for 
distorted closed shell which has been shown to have a substantial contribution to the
nature of D$^{MOI}$. In addition, the angular frequency, $\omega$ associated with
such a motion has been give by,
\begin{equation}
\omega^2=\omega_2^2[1 \pm 4\Omega^2/(\omega_2^2-\omega_3^2)]
\label{eq4}
\end{equation}

\noindent where, $\Omega$ is the angular velocity of the cranking of the nucleus
about a fixed axis. At this juncture, one wonders if such a framework
coupled with the previously mentioned formalism  (Eq.~\ref{eq1}), could be used to directly 
to investigate the moment-of-inertia of the SD bands.  

Interestingly, a simple semi-classical expression between $\Im^{(1)}$ 
and $\Im^{(2)}$ has been derived by C. S. Wu {\it et al.,}
~\cite{wu1987,wu1992,bohr1}. The two parameter expression, using the Bohr Hamiltonian 
for a well deformed nucleus with small axial symmetry has been quite
successful in describing SD band rotational spectrum. Here the authors
have quantified a constant $R$ an a function of the angular momentum $I$, such that 
$R\equiv \sqrt{[\Im^{(1)}]^3/\Im^{(2)}}$. 
One can infer from this formalism, that the moment of inertia can be
represented in a functional form $f_n(\omega)$, $n =1$ and $n=2$, gives,
$\Im^{(1)}$ and $\Im^{(2)}$, respectively. The function is given as,
\begin{equation}
f_n(\omega)=\Im_\circ \left[1-\frac{(\hbar \omega)^2}{a^2b}\right]^{(\frac{1}{2}-n)}
\label{eq5}
\end{equation}
 
\noindent where, $\Im_\circ$, is the band-head moment of inertia, $\Im_\circ$, and 
$a$, and $b$ are related with $\Im_\circ$ via,
\begin{equation}
\Im_\circ=\frac{\hbar ^ 2}{a b}
\label{eq6}
\end{equation} 

The preceding two equations, ~\ref{eq5} and ~\ref{eq6}, indicates that the there is 
slow variation of D$^{MOI}$ with rotational frequency. In addition, the dynamic- and kinematic- 
moment of inertia in SD bands are related to the bandhead moment-of-inertia, 
$\Im_\circ$ through the two parameters, and rotational frequency, $\omega$, as the independent 
variable (Eq.~\ref{eq5}, Eq.~\ref{eq6}).

In our proposed model, the primary motivation of the three semi-classical scenarios 
mentioned previously, have been put together. With this model the 
dynamic-moment-of-inertia, $\Im^{(2)}$, as a function of rotational frequency 
have been investigated. The model is chosen, such that it incorporates a 
vibration-like distortion part and describes the slow variation of D$^{MOI}$ 
for certain SD bands where there is no sudden alignment. 
The function has the property such that, at the bandhead the magnitude of the vibrational
coupling is maximum and it goes to zero at the maximum observed spin $\omega_{max}$.

D$^{MOI}$ as a function of rotational frequency and vibration distortion is given by,
\begin{equation}
\Im^{(2)}= \Im_c^{(2)} \pm \Im_\chi^{(2)} 
\left[ \frac{\omega_{max}-\omega}{\omega_{max}} \right]^2
\label{eq7}
\end{equation}

\noindent where, $\Im_c^{(2)}$, and $\Im_\chi^{(2)}$, are, respectively, the constant 
and the variational part of D$^{MOI}$. Parameter $\Im_\chi^{(2)}$ is such that it 
describes the magnitude of deviation of D$^{MOI}$ from a perfect rotor. 
The quantity $((\omega_{max}-\omega)/ \omega_{max})^2$ is
a function of rotational frequency, $\omega$, such that at $\omega=\omega_{max}$,
it goes to zero. Depending on the initial curvature of the experimental D$^{MOI}$, 
the coupling between $\Im_c^{(2)}$ and $\Im_\chi^{(2)}$ is either positive
or negative. $\Im_\chi^{(2)}$ couples positively to the constant part, if the
vibration is in the plane of rotation. For the negative coupling, the vibration
is more aligned towards the rotation axis. 
 
\section{Formalism}

The atomic nucleus is a finite Fermi system. The value of moment of inertia of 
the nucleus $\Im$, lies between a rigid rotor and a liquid rotor, 
$\Im_{liquid} < \Im <\Im_{rigid}$. One can extract the moment of inertia parameter
for rotational band using the relation, $E_x= \frac{\hbar^2}{2\Im} I(I+1)$. 
However, the average properties of a band in a nucleus 
are better described by the $\Im^{(1)}$, and D$^{MOI}$. 

Experimentally, for most of the SD bands the bandhead spin is not
assigned firmly, due to the missing or low intensity of the intra-band linking transitions.
So, in this work the properties of SD bands have been viewed from the D$^{MOI}$ perspective, and 
the global patterns in the mass region 150 are observed.

The D$^{MOI}$ function is given as in Eq.~\ref{eq7}. The function is single valued and
smooth. The function has a constant part $\Im_c^{(2)}$
and a slowly varying angular frequency dependent part, parametrized through 
$\Im_\chi^{(2)}[ (\omega_{max}-\omega)/ \omega_{max} ]^2 $. Using
least-square fitting the two parameters of this function were extracted from the
experimental D$^{MOI}$ values. Then constant part is related to the MOI of the rigid core
and the variation part is similar to first term of Inglis moment of inertia 
$\Im_{Inglis}$~\cite{Inglis1956}. 

In this model the vibrational distortion effect is parametrized through the effective 
moment-of-inertia, $\Im_\chi$ as a function of rotational frequency. It is noteworthy
that the function is a constant, if $\omega$ is scaled to 
$[(\omega_{max}-\omega)/ \omega_{max} ]^2$. 

\section{Results}
The D$^{MOI}$ (Eq.~\ref{eq7}), have been fitted to the experimental values using 
least square minimum procedure to extract the two parameters, $\Im_c^{(2)}$, and 
$\Im_\chi^{(2)}$. The extracted values are tabulated in Table~\ref{table1}, 
along with the RMS of the residuals for the fit, 
$\sum\limits_{i=1}^n(\Im^{(2)_{cal}}_i-\Im^{(2)_{obs}}_i)^2/n$, 
where n, $\Im^{(2)_{cal}}$, and $\Im^{(2)_{obs}}$ respectively, are the number of 
data points, calculated value of D$^{MOI}$ and observed value of D$^{MOI}$. 
For each of the twenty five cases shown here, the nature of the D$^{MOI}$ in the 
entire rotational frequency range is reproduced quite well.

While, the experimental values of the constant part to D$^{MOI}$, $\Im_c^{(2)}$, 
has a small spread, $72.3^{+ 15.2\%}_{-9.36\%}$, the $\Im_\chi^{(2)}$ 
values span 3.7 $\hbar^2$MeV$^{-1}$ to 88.7 $\hbar^2$MeV$^{-1}$. The characteristics
of D$^{MOI}$ have been identified through the magnitude and coupling (positive or negative)
of the parameter $\Im_\chi$.

\begin{figure*}
\includegraphics[width=0.8\textwidth,natwidth=610,natheight=642]{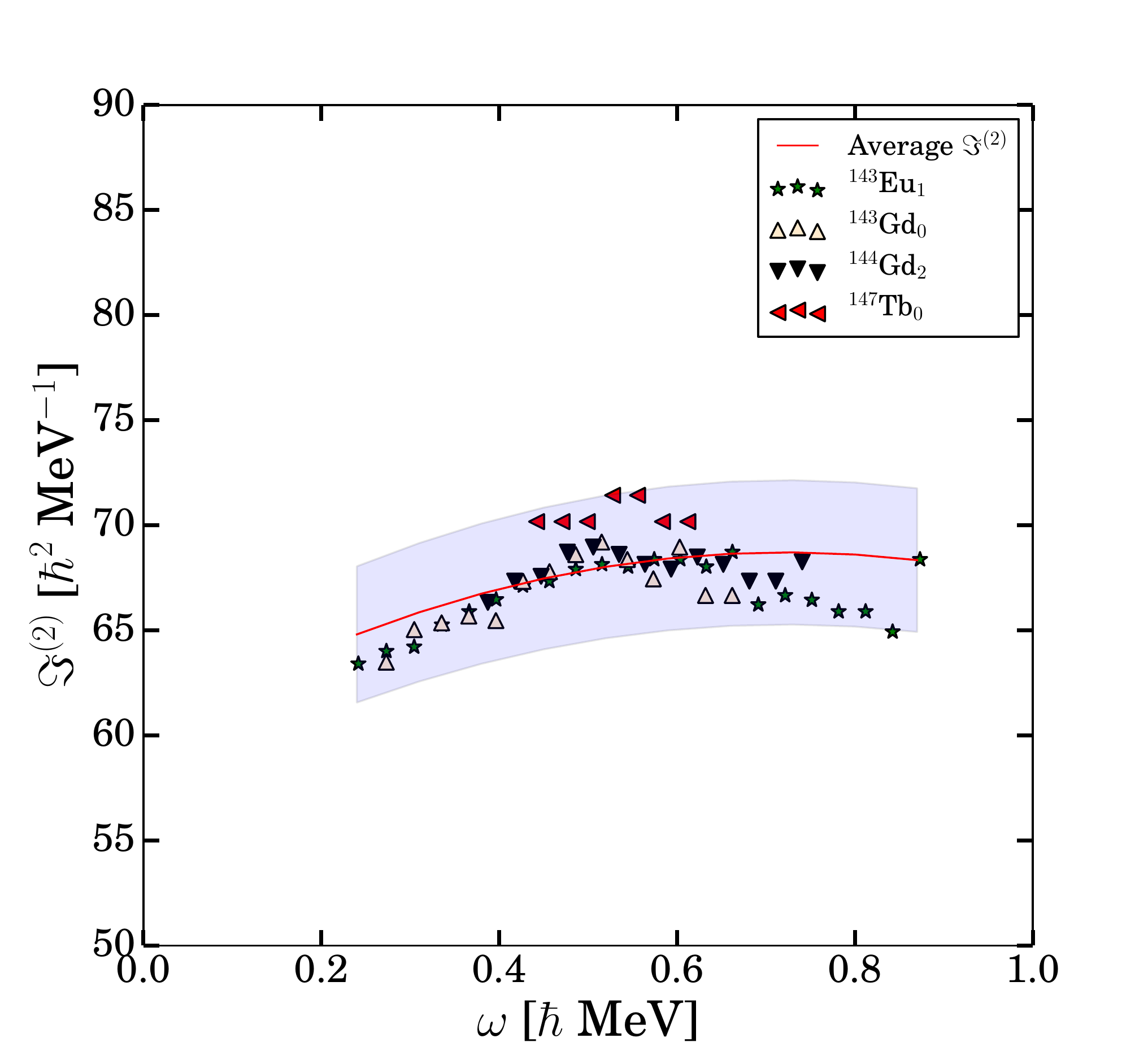}

\caption{Category {\bf I}, $\Im^{(2)}$ vs. $\omega$ for $^{143}$Eu SD band 1, 
$^{143}$Gd SD band, $^{144}$Gd SD band 2 and $^{147}$Tb SD band are shown. 
The red line represents the average curve with the parameter 
value, $\Im^{(2)}_c$ = 68.7 ($\hbar^2$ $MeV^{-1}$) and $\Im_\chi^{(2)}$ = 8.8, respectively.
The blue region spans $\pm 5\%$ of the value of the two parameters. }
\label{fig1}
\end{figure*}

\begin{figure*}
\includegraphics[width=0.8\textwidth,natwidth=610,natheight=642]{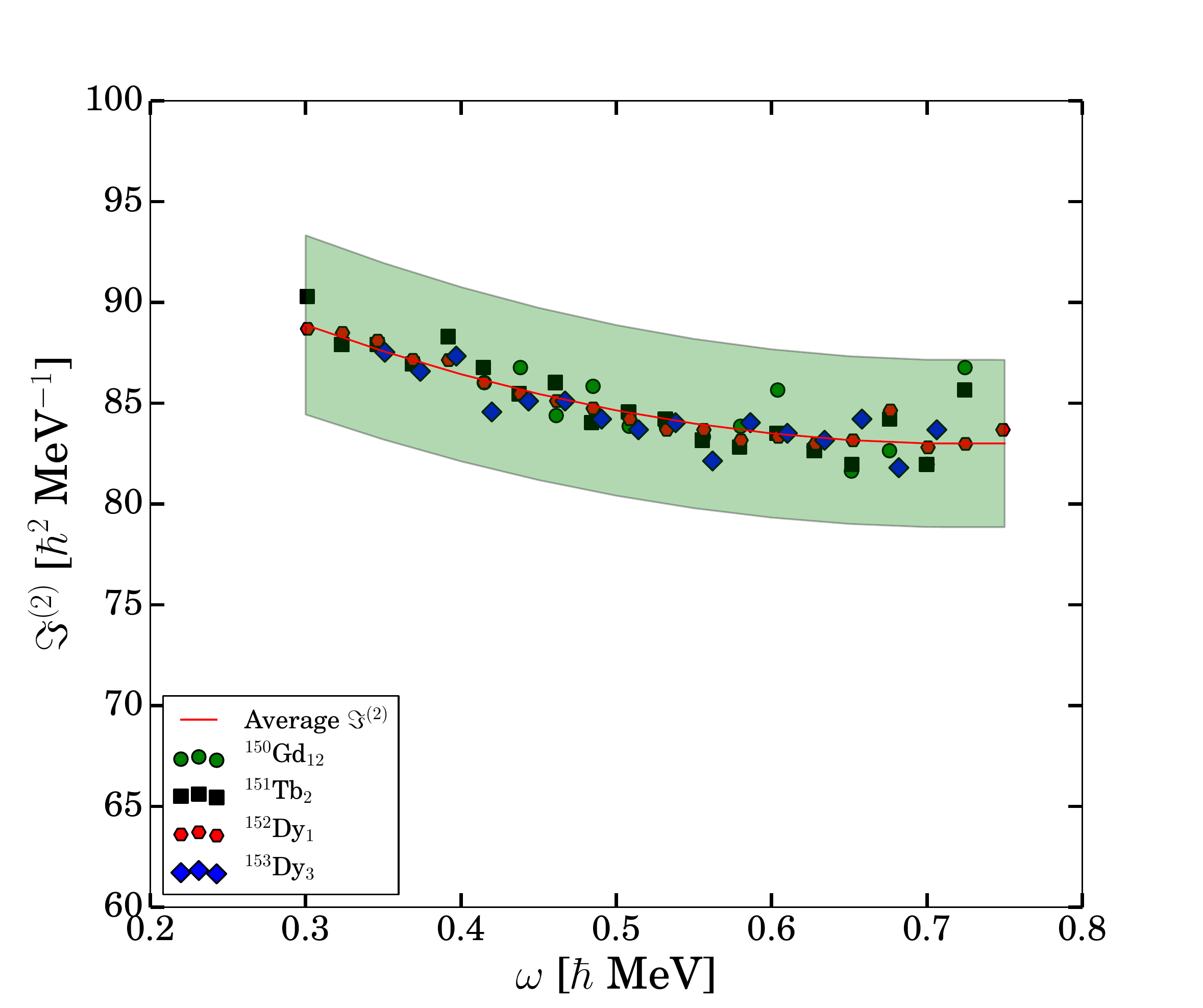}

\caption{Category {\bf III}, $\Im^{(2)}$ vs. $\omega$ for $^{150}$Gd SD band 12, 
$^{151}$Tb SD band 2, $^{152}$Dy SD band 1 and $^{153}$Dy SD band 3 are shown. 
The red line represents the average curve with the parameter value, 
$\Im^{(2)}_c$ = 82.98 ($\hbar^2$ $MeV^{-1}$) and $\Im_\chi^{(2)}$ = 17.13 
($\hbar^2$ $MeV^{-1}$), respectively. The green region spans $\pm 5\%$ of the 
value of the two parameters.}
\label{fig2}
\end{figure*}

\begin{figure*}
\includegraphics[width=0.8\textwidth,natwidth=610,natheight=642]{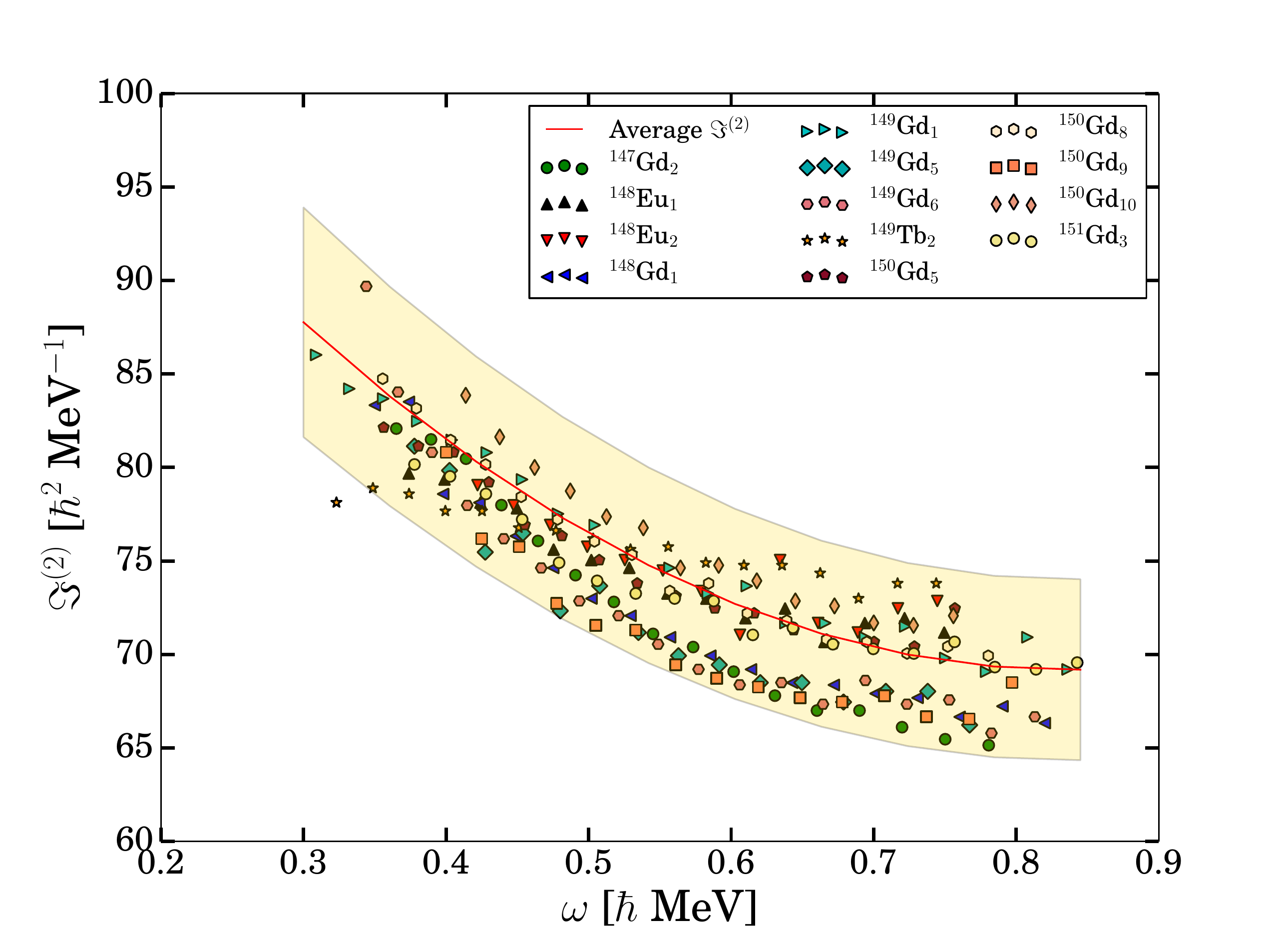}

\caption{Category {\bf III}, $\Im^{(2)}$ vs. $\omega$ for $^{147}$Gd SD band 2, $^{148}$Eu SD band 1,
$^{148}$Eu SD band 2, $^{148}$Gd SD band 1, $^{149}$Gd SD band 1,$^{149}$Gd SD band 5,
$^{149}$Gd SD band 6, $^{149}$Tb SD band 2, $^{150}$Gd SD band 5, $^{150}$Gd SD band 8,
$^{150}$Gd SD band 9, $^{150}$Gd SD band 10, and $^{151}$Gd SD band 3. 
The red line represents the average curve with the parameter value, 
$\Im^{(2)}_c$ = 69.18  ($\hbar^2$ $MeV^{-1}$) and $\Im_\chi^{(2)}$ = 45.21 
($\hbar^2$ $MeV^{-1}$), respectively. The yellow region spans $\pm 7\%$ of the 
value of the two parameters. }
\label{fig3}
\end{figure*}

In the figures~\ref{fig1}, ~\ref{fig2}, and ~\ref{fig3}, the nuclei with three distinct
patterns in D$^{MOI}$ components are plotted. Three categories of D$^{MOI}$, {\bf I}, 
{\bf II}, and {\bf III}, are identified depending on the magnitude of the ratio of the 
D$^{MOI}$ component, $\Im_c^{(2)}$ and $\Im_\chi^{(2)}$ and nature of the coupling. 
In the category {\bf I}, the average magnitude of $\Im_c^{(2)}$ is about eight times
the average magnitude of $\Im_\chi^{(2)}$ (Fig.~\ref{fig1}). The nuclei,
$^{143}$Eu, band 1, $^{143}$Gd yrast SD band, $^{144}$Gd band 2, and, 
$^{147}$Tb yrast SD band, fall under this category. In this region two other
SD bands, $^{150}$Tb SD band 2, and $^{151}$Dy SD band 1 also show similar
magnitude of the negative coupling of $\Im_\chi^{(2)}$, but the magnitude of 
$\Im_c^{(2)}$ is quite higher and hence not included in the category {\bf I}. 
 
Thus, the region {\bf I} is characterized by the averages $\Im_c^{(2)}$ = 
68.7 ($\hbar^2$ $MeV^{-1}$)), and $\Im_\chi$ = 8.8 ($\hbar^2$ $MeV^{-1}$). 
Curiously, the coupling between the D$^{MOI}$ component is negative, hence, 
the initial D$^{MOI}$ starts with a lower value, that gradually increases 
and becomes equal to $\Im_c$ at $\omega=\omega_{max}$. The shaded blue region 
in the figure ~\ref{fig1} corresponds to $\pm 5 \%$ of the average values, 
depicted by the red line. All the experimental $\Im^{(2)}$ values are 
found to be within the region. 

In the category {\bf II} (Fig.~\ref{fig2}), the coupling between $\Im_c^{(2)}$ 
and $\Im_\chi^{(2)}$ is positive, and average value of the D$^{MOI}$ components are 
$\Im_c^{(2)}$ = 82.98 ($\hbar^2$ $MeV^{-1}$) , and $\Im_\chi^{(2)}$ = 17.1 
($\hbar^2$ $MeV^{-1}$). Here, $\Im_c^{(2)}$ is higher than other two region and about 
five times the average $\Im_\chi^{(2)}$ value. As a result
of the coupling the initial D$^{MOI}$ is higher and decreases to $\Im_c^{(2)}$ value 
at $\omega=\omega_{max}$. $^{150}$Gd SD band 12, $^{151}$Tb SD band 2,
$^{152}$Dy SD band 1, and $^{153}$Dy SD band 3 belong to this category. 
The green region represents the $\pm 5\%$ of the average values shown
by the red line. 

In the final category, {\bf III}, we have thirteen bands, where $\Im_c^{(2)}$ is about
1.5 times $\Im_\chi^{(2)}$ and the coupling is positive like the previous category. 
Due to considerable increase in the magnitude of the frequency dependent part, 
the initial D$^{MOI}$ is high and decreases much rapidly than the other two categories. 
The  D$^{MOI}$, components $\Im_c^{(2)}$ and $\Im_\chi^{(2)}$ have comparable magnitude and on 
the average are given as, 69.3 ($\hbar^2$ $MeV^{-1}$) and 47 ($\hbar^2$ $MeV^{-1}$), respectively.

\section{Summary}

A empirical semi-classical two parameter model have been used to investigate the 
characteristics of D$^{MOI}$ of SD band in mass 150 region. Using the model 
three categories of D$^{MOI}$ have been identified. The D$^{MOI}$ in this model 
have been parametrized such that, the component $\Im_c^{(2)}$ is constant
and $\Im_\chi^{(2)}$ is dependent on angular frequency. Experimental values, 
of twenty five SD bands, which show a smooth behaviour have been fitted 
using the equation ~\ref{eq7}. In all the cases we have a reasonably good fit.

\begin{figure*}
\includegraphics[width=0.8\textwidth,natwidth=610,natheight=642]{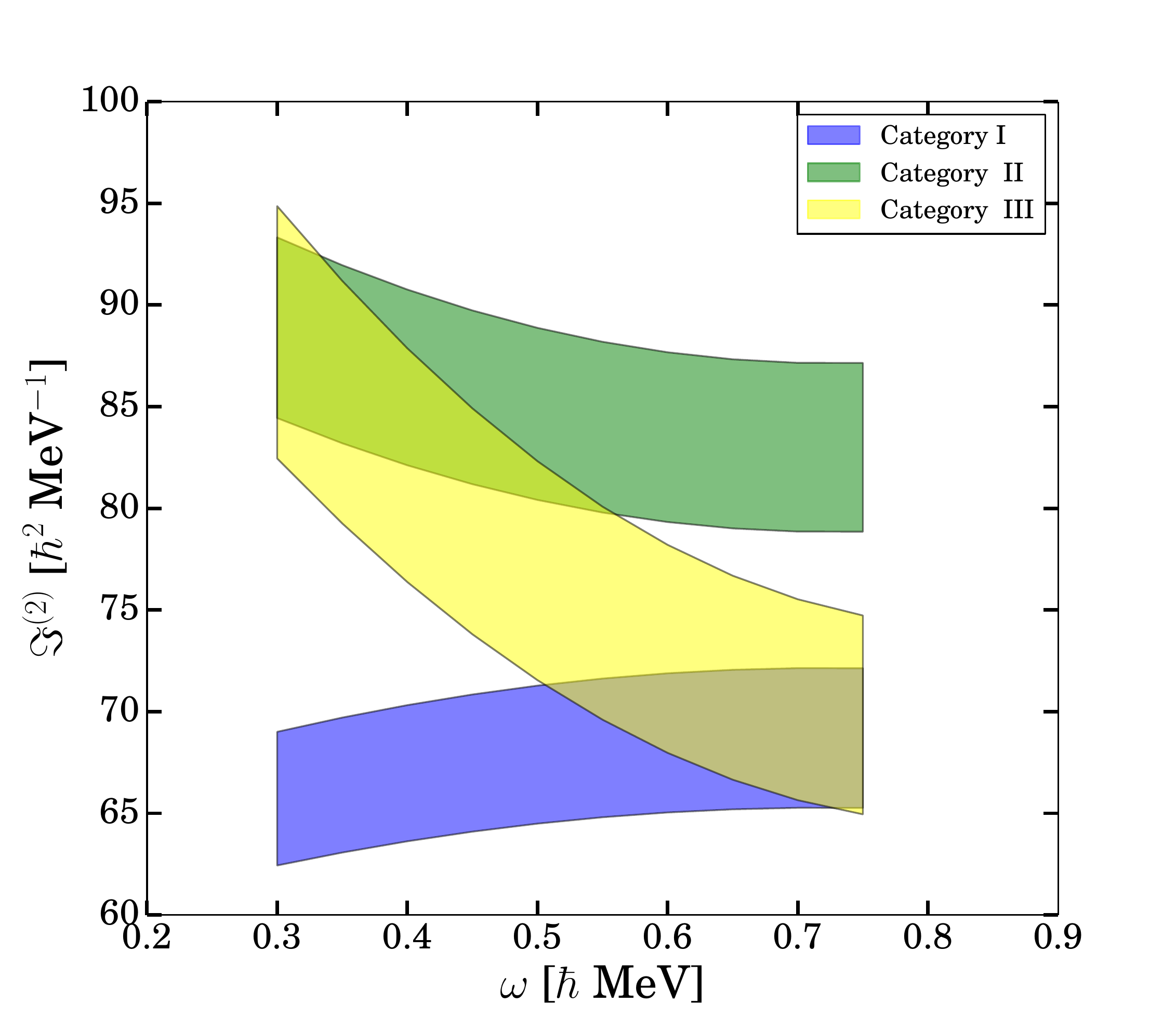}

\caption{The $\Im^{(2)}$ vs. $\omega$ plot shows the three regions with 
$\Im^{(2)}_c/\Im^{(2)}_\chi$ $\sim 8$ (blue), $\Im^{(2)}_c/\Im^{(2)}_\chi$ $\sim 5$ (green),
and $\Im^{(2)}_c/\Im^{(2)}_\chi$ $\sim 1.5 $ (yellow) for the given span of
angular frequency.
 }
\label{fig4}
\end{figure*}

The three categories of the D$^{MOI}$, are shown for a given angular frequency 
range figure~\ref{fig4}. It is interesting to note that at the lower frequency 
limit, category {\bf II} and category {\bf III}, D$^{MOI}$ overlap. 
This indicates that in both the scenario, the bandhead D$^{MOI}$ is similar, but
the coupling of the angular frequency dependent component is different. 
At higher angular frequency, these two DMOI become distinctly separate and 
reach the respective average rigid core values of 82.98 ($\hbar^2$ $MeV^{-1}$) 
and 69.3 ($\hbar^2$ $MeV^{-1}$) for category {\bf III} and category {\bf II}.

At the higher frequency limit, D$^{MOI}$ category {\bf I} and {\bf II} overlap. 
The coupling of $\Im_{\chi}^{(2)}$ is negative and this results in a lower
D$^{MOI}$ at the bandhead. At the higher frequency the this effect diminishes
and two category of D$^{MOI}$ {\bf II} and {\bf I} merges to the average value
of $\sim$ 69 ($\hbar^2$ $MeV^{-1}$).

In this study three distinct patterns of the D$^{MOI}$ have been highlighted.  
A substantial number of SD bands have been treated using the model and good
agreement is observed between the experiment and theory. As a result, the D$^{MOI}$ 
may be thought to have a rather weak dependence on their respective nucleonic 
configurations. The angular frequency component given by, 
$[(\omega_{max}-\omega)/\omega_{max}]^2$, is a direct measure of the
distortion effect and responsible for the deviation from the ideal rotor. It is
natural that this model will also work for smooth SD bands in other mass region.
The present model can also be extended to calculate the B(E2) rates, by parametrizing 
the charge asymmetry through the distortion effect. 

\section{Acknowledgement}

Discussions with R. G. Pillay have been very constructive in the preparation 
of this manuscript and his insight on the subject matter has been very helpful.

\newpage
%---------------

%---------------
%---------------

%---------------

\newpage
\begin{table}
\centering
\caption{\label{tab:table1}~ Fitted parameter values, $\Im_\circ^{(2)}$, $\Im\chi^{(2)}$
using the Eq.~\ref{eq7} are tabulated.}
\begin{adjustbox}{max width=\textwidth}
\begin{ruledtabular}
\begin{tabular}{|c|c|c|c|}
\hline
Isotope(Band) & $\Im_\circ^{(2)}$ & $\Im\chi^{(2)}$ & rms of residuals \\
\hline
$^{143}$Eu(SD1)&67.4882&5.51486&1.30959\\
$^{143}$Gd(SD)&68.2859&12.0957&1.00398\\
$^{144}$Gd(SD2)&68.2249&3.6832&0.703386\\
%$^{145}$Gd(SD1)&70.4495&48.8543&1.0384\\
$^{147}$Gd(SD2)&65.6736&62.4044&0.515283\\
$^{147}$Tb(SD)&70.8534&13.7481&0.638003\\
$^{148}$Eu(SD1)&71.209&35.9854&0.521523\\
$^{148}$Eu(SD2)&71.8992&37.1367&1.07683\\
$^{148}$Gd(SD1)&66.3205&50.523&0.817359\\
$^{148}$Gd(SD6)&70.5666&70.2551&0.617295\\
$^{149}$Gd(SD1)&69.9208&41.7947&0.621519\\
$^{149}$Gd(SD5)&66.8639&52.4335&0.99385\\
$^{149}$Gd(SD6)&65.5212&58.6527&1.71442\\
$^{149}$Tb(SD2)&74.1203&16.4275&0.546993\\
%$^{150}$Gd(SD1)&74.9729&152.198&3.0639\\
$^{150}$Gd(SD5)&70.6088&42.7737&0.667379\\
$^{150}$Gd(SD8)&70.0014&49.3173&0.307386\\
$^{150}$Gd(SD9)&66.3469&47.9009&1.18063\\
$^{150}$Gd(SD10)&71.6982&56.9163&0.37468\\
$^{150}$Gd(SD12)&83.2367&15.7484&1.48063\\
$^{150}$Tb(SD1)&73.4802&17.9999&0.57067\\
$^{150}$Tb(SD2)&77.2242&13.5021&1.26347\\
$^{151}$Dy(SD1)&79.5679&7.71123&1.51678\\
%$^{151}$Gd(SD1)&73.2077&88.6882&2.00907\\
%$^{151}$Gd(SD2)&73.3791&82.1452&2.09608\\
$^{151}$Gd(SD3)&69.2381&35.52&0.594611\\
$^{151}$Tb(SD2)&82.8791&19.2366&1.03616\\
$^{152}$Dy(SD1)&82.906&16.4917&0.523973\\
$^{153}$Dy(SD3)&82.91&17.0606&0.812715\\
\hline
\end{tabular}
\end{ruledtabular}
\end{adjustbox}
\label{table1}
\end{table} 
\end{document}